# EVOLUTION OF FEEDBACK LOOPS IN OSCILLATORY SYSTEMS

M. Hafner[1,2,3], H. Koeppl[1,4] and A. Wagner[2,3,5,6]

[1] School of Computer and Communication Sciences, Ecole Polytechnique Fédérale de Lausanne (EPFL), CH-1015 Lausanne, Switzerland
[2] Department of Biochemistry, University of Zurich, CH-8057 Zurich, Switzerland
[3] Swiss Institute of Bioinformatics, CH-1015 Lausanne, Switzerland
[4] Plectix Biosystems, Somerville, MA 02144, USA
[5] The Santa Fe Institute, Santa Fe, NM 87501, USA
[6] Department of Biology, University of New Mexico, Albuquerque, NM 87131, USA

*Abstract*

Feedback loops are major components of biochemical systems. Many systems show multiple such (positive or negative) feedback loops. Nevertheless, very few quantitative analyses address the question how such multiple feedback loops evolved. Based on published models from the mitotic cycle in embryogenesis, we build a few case studies. Using a simple core architecture (transcription, phosphorylation and degradation), we define oscillatory models having either one positive feedback or one negative feedback, or both loops. With these models, we address the following questions about evolvability: could a system evolve from a simple model to a more complex one with a continuous transition in the parameter space? How do new feedback loops emerge without disrupting the proper function of the system? Our results show that progressive formation of a second feedback loop is possible without disturbing existing oscillatory behavior. For this process, the parameters of the system have to change during evolution to maintain predefined properties of oscillations like period and amplitude.

**Introduction**

Feedback loops are considered key motifs of biochemical systems such as signaling pathways, homeostatic regulatory circuits and oscillators (Alon, 2007). In signaling pathways, different wirings of feedback loops result in a broad spectrum of signal responses (Tyson et al., 2003). In oscillatory systems, one feedback component is essential to create oscillations (Goldebeter, 1996). In homeostatic control circuits these loops allow an efficient control of the concentrations of a system's different species (El-Samad et al., 2005). Multiple feedback loops can also result in a better robustness of such systems (Hastings, 2000).

From an evolutionary point of view, it is not clear how multiple loops evolved in complex systems like the early embryonic cell cycle (Pomenenring et al., 2003) or in circadian clocks (Wijnen and Young, 2006). In silico approaches have been used to understand how a system with a transcriptional repression pattern can evolve oscillations (François and Hakim, 2004), and if such systems can show oscillations by tuning their kinetics parameters (François and Hakim, 2005 and Fung et al., 2005). On the scale of the network topology (neglecting kinetic parameters), oscillations can be conserved while modifying the structure of a network (Wagner, 2005).

All these studies do not address the question whether new feedback loops can be created in an oscillatory system without perturbing existing oscillatory behavior. Even if evolution occurs in finite steps, it seems very unlikely a priori that a new loop can be created at random with precisely the correct parameters to maintain the existing period and sufficiently high amplitude of oscillation. The emergence of a new feedback loop would more probably occur in multiple small steps which facilitate adjustment of kinetics parameter to maintain core oscillatory properties. In this paper, we will show precisely that: evolution of this kind is possible for simple models that have been used to model the mitotic cell cycle (Pomenenring et al., 2003). We propose two models based on early models from the mitotic cycle in embryogenesis (Goldbeter, 1996 and Tyson et al., 2003). These models have one feedback loop, either positive or negative. Both show oscillations for a broad range of kinetics parameters. We evolve these models toward a system with both positive and negative feedback loops.

Using a Monte Carlo approach, we will show that a simple model can evolve toward a more complex one with small changes in the value of its kinetics parameters. We will find evolutionary paths in the parameter space that conserve the *viable properties* of the system. In the considered oscillatory models, such conservation means that period and amplitude of oscillation have to stay within predetermined intervals.

**Methods**

*Random Walk for Evolution*

We first describe an algorithm to study the emergence of new regulatory motifs in a model. Starting from a model, say **M₁**, with a nominal parameter set, the goal is to evolve through small changes of the parameters toward a predefined second model, say **M₂** that is similar to **M₁**, but has an additional motif. During evolution, some *viable properties* of the system have to be conserved. For example, in a homeostatic mechanism, the target concentration of a molecule may have to stay in a specific interval; in this case evolution of a new motif could allow faster or more reactive control (El-Samad et al., 2005). For oscillatory systems, the period and the amplitude have to be conserved while better robustness can appear through new motifs (Hastings, 2000).

In practice, the parameters of the new motif in model **M₂** are set to zero in order to mimic the nominal parameter set of **M₁**. The aim of the evolution process is to change progressively these parameters, and if necessary other parameters of the model, in order to reach their nominal values in model **M₂**. We allow a maximum of 10% variation in the parameter space at each step to mimic the fact that most mutations may affect biochemical parameters to a small extent.

Our algorithm starts with a biased random walk in the logarithmic domain (Eq. (1)).

$$\vec{\kappa}^{j+1} = \vec{\kappa}^{j} + \alpha\ \varepsilon + \beta\ \vec{\Delta}^{(j)} / \|\vec{\Delta}^{(j)}\| \qquad (1)$$

Where $\vec{\kappa}^{j+1} = \log_{10}(k_1), \log_{10}(k_2), ...^{(j+1)}$ is the logarithm of the parameter set at the *j*-th iteration of the random walk, $\vec{\Delta}^{(j)} = \log_{10} \vec{k}^{(M_2)} - \log_{10} \vec{k}^{(j)}$ is the difference between the *j*-th parameter set and the nominal set for model **M₂** in the logarithmic domain. The random vector $\varepsilon$ is normally distributed with independent components. The value α = 0.041 is chosen such that the standard deviation of parameter variation is 10% of the previous parameter value; we set β to a value of 1/3.

This random walk finds *viable* points in the parameter space, i.e. parameter sets for which a model shows predefined viable properties. When the random walk has reached the vicinity of the nominal set for **M₂**, we shorten the path by reducing the number of points by linear interpolation between distant points along the path. As we choose a priori the maximal length for the line segments, we do not recover shortest possible path. During this process, we check that the viable properties are conserved along the line connecting the two intermediate points. If the whole path consists of viable points and the connections between them are also viable, we consider the properties to be conserved along the evolution process.

*Models*

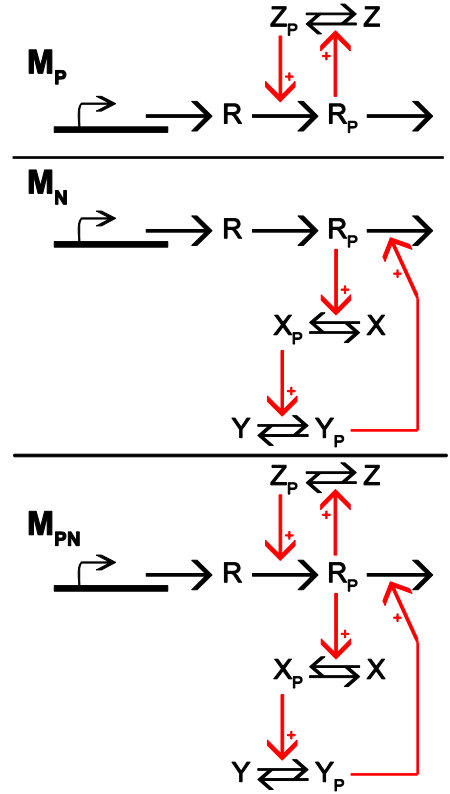

*Figure 1: Reaction diagrams of the three models. On the top, $M_P$ with only the positive feedback loop. In the middle, $M_N$ with only the negative feedback loop. On the bottom, $M_{PN}$ with both feedback loops.*

The mitotic cycle has been one of the first biological systems to be modeled with feedback loops. The first two published models were based either on a positive feedback (Tyson, 1991) or a negative one (Goldbeter, 1991). Further models were published including both kinds of loops (see Tyson et al., 2003, Pomenenring et al., 2003 and ref. therein). The models for our case study are inspired by models in these papers, because they are simple yet biologically realistic. Specifically, we propose three models with different feedback architectures, shown in Figure 1. All our models are based on the expression of a protein *R*, its phosphorylation and its degradation. For the model with positive feedback (**M_P**), the phosphorylated state of the protein ($R_P$) acts as a kinase for a secondary protein $Z$ ($Z \leftrightarrow Z_P$). The positive feedback loop is formed by $Z_P$ that increases the rate of the reaction $R \rightarrow R_P$. For the model with negative feedback ($\mathbf{M_N}$), an intermediate step is needed to introduce more delay: $R_P$ acts as a kinase for an intermediate protein $X$ ($X \leftrightarrow X_P$) and $X_P$ phosphorylates a third protein $Y$ ($Y \leftrightarrow Y_P$). The phosphorylated state of this protein, $Y_P$, increases the degradation rate of $R$, therefore acting as a negative feedback. For the more complex model ($\mathbf{M_{PN}}$), both loops are present: $R_P$ influences the phosphorylation of $X$ and $Z$. To translate the models into differential equations, we assume that the reactions involving $R$ and $R_P$ follow mass-action kinetics:

$$\frac{d[R]}{dt} = k_1 - p\,[Z_P]\,[R]$$
$$\frac{d[R_P]}{dt} = p\,[Z_P]\,[R] - n\,[Y_P]\,[R_P] \quad (2)$$

The functions $p([Z_P])$ and $n([Y_P])$ reflect the positive and negative feedbacks contained in model $\mathbf{M_{PN}}$:

$$p\,[Z_P] = k_2 + k_{11}[Z_P]$$
$$n\,[Y_P] = k_3 + k_{12}[Y_P] \quad (3)$$

In models $\mathbf{M_P}$ and $\mathbf{M_N}$, $k_{12}$, resp. $k_{11}$, are set to zero such that only one feedback is effective. The phosphorylation of $X$, $Y$, and $Z$ are governed by Michaelis-Menten kinetics:

$$\frac{d[X_P]}{dt} = \frac{k_4[R_P]\,[X_T]-[X_P]}{k_{10}+[X_T]-[X_P]} - \frac{k_5[X_P]}{k_{10}+[X_P]}$$
$$\frac{d[Y_P]}{dt} = \frac{k_6[X_P]\,[Y_T]-[Y_P]}{k_{10}+[Y_T]-[Y_P]} - \frac{k_7[Y_P]}{k_{10}+[Y_P]} \quad (4)$$
$$\frac{d[Z_P]}{dt} = \frac{k_8[R_P]\,[Z_T]-[Z_P]}{k_{10}+[Z_T]-[Z_P]} - \frac{k_9[Z_P]}{k_{10}+[Z_P]}$$

The terms $[X_T]$, $[Y_T]$ and $[Z_T]$ denote the total concentration of proteins $X$, $Y$ and $Z$, respectively. We choose them to be equal to 1, and note that the absolute value is irrelevant, because a proper scaling of the parameters allows changing time and concentration independently. In order to simplify the notation, we will not write the units for time, concentration and parameters.

*Viable Parameter Sets*

With the above-defined random walk in the 12-dimensional parameter space of our models we can identify new viable parameter sets. The chosen viable properties of our models are related to the oscillations of the concentration of $R_P$. In particular, we want the values of the period, the peak value and the amplitude of these oscillations to remain within predetermined intervals. A parameter set is called viable if the concentration of $R_P$ oscillates with a period in the (arbitrary) interval [0.9, 1.1], a peak value contained in the range [0.5, 1.0] and an amplitude that is at least 40% of the peak value. We chose these criteria to reflect an important feature of biological oscillations, namely that they avoid very small amplitudes. Period and the peak values are readily adjusted with a proper scaling of the parameters.

For models $\mathbf{M_P}$ and $\mathbf{M_N}$, the nominal parameter sets were taken from Tyson et al. (2003) and rescaled in order to have a period of 1 and a peak value of 0.66. For the model $\mathbf{M_{PN}}$, the nominal parameters are chosen to be of the same order of magnitude as for models $\mathbf{M_P}$ and $\mathbf{M_N}$ (in particular they were obtained as the average of both sets and then rescaled to fulfill the same viable properties).

**Results**

For most of our simulations (>85%) our algorithm is able to connect one of the one-loop models to the two-loop model (refer to Figure 2 and Figure 3). Usually about a few thousands of points are tested during the random walk, half of which are viable. The path is then reduced to about one hundred points connected with viable segments.

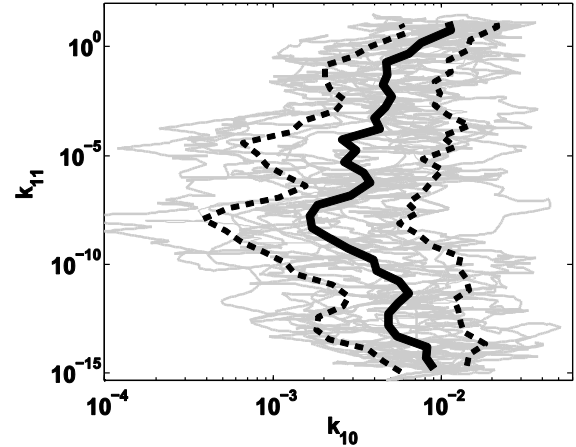

*Figure 2: 15 different paths from model $M_N$ to model $M_{PN}$ projected on the plane ($k_{10}$, $k_{11}$). Parameter $k_{11}$ is increased during evolution. Light gray lines correspond to paths; the black line is the average path, and dashed black lines are the standard deviations along the average path.*

*$M_N$ to $M_{PN}$ Evolution*

The addition of a positive feedback loop to the model with only a negative feedback loop is also possible, but the straight line connecting the two parameter sets for model $M_N$ and $M_{PN}$ is not viable. The line crosses a Hopf-bifurcation, where the amplitude decreases and then oscillations disappear. Therefore to connect both models, the random walks follow a bent trajectory. The most significant adaptation is seen for the Michaelis constant, $k_{10}$ (Fig. 2). When $k_{10}$ decreases, the transitions $(X \leftrightarrow X_P)$ and $(Y \leftrightarrow Y_P)$ become switch-like which strengthens the nonlinearity and may support oscillatory behavior.

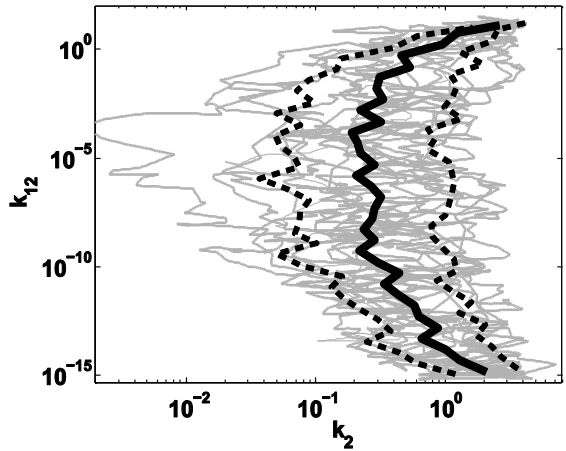

*Figure 3: 15 different paths from model $M_P$ to model $M_{PN}$ projected on the plane $(k_2, k_{12})$. Parameter $k_{12}$ is increased during evolution. Light gray curves correspond to paths; the black line is the average path, and dashed black lines are the standard deviations along the average path.*

### $M_P$ to $M_{PN}$ Evolution

The addition of a new negative feedback loop to the model with only one positive feedback loop is also possible. In this case also, the straight line in the parameter space is not viable because the period increases beyond the allowed maximal value along the path. The random walks become biased toward smaller values of the phosphorylation rate of $R$, $k_2$ (Fig. 3). This can be explained by the fact that $k_2$ has a positive impact on the period. Decreasing it also decreases the period.

### Conclusion

This work is a first step to understand the emergence of feedback loops in oscillatory system. We have shown that such evolution is possible with many small adaptations of the kinetic parameters. For our two case studies, the addition of a second loop is possible only if other parameters are changed during this process.

In such systems, the high-dimensionality prohibits classical qualitative bifurcation analyzes. Moreover, if quantitative constraints on system function have to be fulfilled, the problem cannot be solved analytically. The random sampling approach is limited by the computational cost, which is very high for large ranges of parameters in high dimensions. In this case, the random walk approach works well and is a promising tool to understand the evolution of more complex systems, such as the mammalian circadian clock.

### Acknowledgments

All authors are grateful to SystemsX.ch grant for IPhD Project "Quantifying Robustness of Biochemical Modules to Parametric and Structural Perturbations". HK acknowledges the support from the Swiss National Science Foundation (SNF), grant 200020-117975/1. AW acknowledges support from SNF grants 315200-116814 and 315200-119697, as well as from the SystemsX.ch RTD project "Coping with uncertainty: Towards an integrated understanding of nutrient signaling, regulation and metabolic operation".